\documentclass[runningheads]{llncs}
\usepackage[T1]{fontenc}
\usepackage{graphicx}
\usepackage{amsmath}
\begin{document}
%
%
%
%
\title{Words to Waves: Emotion-Adaptive Music
Recommendation System}
\author{Apoorva Chavali\textsuperscript{1} \and Reeve Menezes\textsuperscript{2}}

\institute{Virginia Tech}
\maketitle 
\begin{abstract}

Current recommendation systems often tend to overlook emotional context and rely on historical listening patterns or static mood tags. This paper introduces a novel music recommendation framework employing a variant of Wide and Deep Learning architecture that takes in real-time emotional states inferred directly from natural language as inputs and recommends songs that closely portray the mood. The system captures emotional contexts from user-provided textual descriptions by using transformer-based embeddings, which were finetuned to predict the emotional dimensions of valence-arousal. The deep component of the architecture utilizes these embeddings to generalize unseen emotional patterns, while the wide component effectively memorizes user-emotion and emotion-genre associations through cross-product features. Experimental results show that personalized music selections positively influence the user's emotions and lead to a significant improvement in emotional relevance.

\keywords{Recommendation systems  \and embeddings \and Wide and Deep learning \and valence \and arousal}
\end{abstract}
\section{Introduction}

With the rapid advancement of artificial intelligence, many music recommendation systems using machine learning models are emerging in today’s entertainment market. Music is inherently emotional and can be expressed through various lyrics, genres, and sounds. Emotional responses to music can trigger pleasure and reward regions of the brain, establishing a link between auditory and emotional arousal \cite{BloodZatorre2001}. Numerous music databases and libraries exist under various categories to encompass diverse user emotions. However, a significant challenge in this industry is matching a user’s emotional state to the song being played, as human emotions are unstable and cannot be captured with just a few static tags, which lack variability \cite{Grewal2021}\cite{Brinker2012}.
Most major music application companies rely on conventional recommender systems that predominantly use historical listening patterns or fixed mood classifications. Recent works show that graph recommendation systems are being built where the essential objects, e.g., users, items, and attributes, are either explicitly or implicitly connected \cite{graphrec}. These systems frequently neglect listeners' dynamic and real-time emotional states, leading to recommendations that may feel irrelevant or disconnected from the user’s current emotional context. 

To address this issue, we propose an advanced music recommendation framework that utilizes a Wide and Deep Learning architecture with the capabilities of both generalization and memorization, which is essential for a recommendation system. The user's natural language inputs are mapped onto a valence-arousal (VA) space using a transformer-based approach \cite{MERTransformer2024}\cite{AudioEmotion2025}, where valence (V) indicates how calm or agitated a user feels, and arousal (A) reflects the positivity or negativity of their state. By quantifying these metrics, we can infer a user’s emotional state and recommend a song that closely matches that feeling.

With this new system, users no longer need to scroll endlessly to find the right song that aligns with their current emotion, but the system instantly recommends music, which adds an element of surprise and discovery, as users are not constrained to selecting a specific song. This system also aids in the exploration of new artists and genres, giving them more reach. At its core, our system shifts the focus from “What did you like before?” to the more immediate and emotionally relevant question: “How do you feel now?”

We utilize wide and deep architecture \cite{inproceedings} as it combines the power of shallow and deep learning techniques, which have been proven to improve user recommendations significantly. The wide component is designed for memorization, and it captures patterns and associations such as user-emotion and emotion-artist preferences. It uses structured features like sentence length buckets (to capture structural patterns) in an emotional context to support known relationships. The deep component is used to generalize various user expressions. It maps the natural language input to a valence-arousal space by identifying emotional patterns that may not be explicitly seen during training. Thus, this structure strikes a balance between generalization and memorization, achieving robust performance on both familiar and novel emotional queries. It is to be noted that although our architecture keeps the two branch idea of the traditional wide and deep architecture intact, adding hidden layers to the wide side and merging the branches by concatenation (instead of a simple add) makes it only a Wide-and-Deep-inspired variant, not the original architecture.

\section{Related Work}

Recommendation systems' applications are developing rapidly with the increase in content and users across the internet. Traditional collaborative filtering and content-based systems have been extensively used by commercial platforms such as Spotify and Pandora. These models use user-item interactions and metadata to infer user preferences, yet do not consider real-time emotions. Hybrid models combining content and collaborative signals have been proposed to enhance personalization, but still lack sensitivity to users' current emotional states. Roy et al. \cite{s40537-022-00592-5} present a comprehensive survey of recommender systems, analyzing 60 peer-reviewed articles from 2011 to 2021. It also talks about problems like difficulty in recommending items to new users, or recommending new items, and insufficient user-item interactions leading to less accurate recommendations. It also mentions that combining content-based filtering (e.g., song features) with collaborative filtering \cite{collaborative-filtering} (e.g., user listening history) can enhance recommendation accuracy. This has been implemented by Google in their paper \cite{inproceedings} where they introduce a deep and wide architecture. They combine the strengths of memorization (wide component) and generalization (deep component) for recommender systems. The wide part captures explicit feature interactions, while the deep part generalizes to unseen feature combinations through embeddings. This system is implemented in Google Play to increase app acquisitions and in YouTube to improve video recommendations. 
Ayata et al. \cite{ayata} proposed a music recommendation model based on the user emotions captured through wearable physiological sensors like Galvanic Skin Response and Photo Plethysmography Signals. They used the emotion valence-arousal dimensional model to determine the user's emotions. The emotion detection algorithm employed different machine learning algorithms like SVM, RF, KNN, and decision tree algorithms to predict the emotions from the changing electrical signals gathered from the wearable sensors. Antoine Toisoul et al. \cite{va-paper} estimated the valence and arousal values from facial images.

Erkang Jin et al. \cite{3} address the variability in users' emotional expressions and music preferences by introducing a Heterogeneity-aware Deep Bayesian Network. It accounts for differences in emotion perception across users and within the same user over time. Their model can adapt to emotional heterogeneity (i.e., that users react differently to the same emotion category), but it doesn’t process free-text emotion descriptions dynamically. Kin Wai Cheuk \cite{4} used triplet neural networks for regression tasks to predict music emotions in terms of valence and arousal. Tkalčič \cite{5} presented a framework for emotion-aware recommender systems, using the valence-arousal-dominance (VAD) model to describe users' emotional reactions to content. The study includes a case study on image recommendation, demonstrating how incorporating emotional metadata can enhance recommendation performance. Li et al. \cite{beamers} utilized EEG data to predict users' emotional variations in the valence-arousal model. This research proves the importance of valence-arousal space in modulating users' emotional state. 

A few researchers used deep convolutional neural networks with weighted feature extraction to correlate user data with music features for emotion-based recommendations \cite{app8071103}. The model maps users to fixed emotion categories (e.g., happy, sad) rather than handling dynamic expressions. CNNs are suitable for feature extraction from structured inputs (e.g., spectrograms or fixed-length vectors), but they underperform on variable-length natural language where semantics and context matter. 

There is an interdependence of words in a sentence that captures emotions better than tags like happy, sad, anxious, etc. Humans have always tried to put down emotions in poems and literature, which capture mixed emotions. Deep et al. \cite{sentence_imp} talk about how sentence context is crucial for accurate emotion recognition. Singh et al.  \cite{sentence_imp2} adapts a language model to generate emotionally expressive text, demonstrating that sentences can be crafted to convey specific emotions with controlled intensity, and this adaptability showcases the role of sentence construction in emotional expression.

\section{Approach}
\subsection{Datasets}
We utilized three datasets for three different purposes:

1. EmoBank \cite{buechel-hahn-2017-emobank}: A corpus of 10,000 English sentences with multiple genres, which were annotated with dimensional emotion metadata in the VAD representation format. Each sentence in the dataset is labeled by a human with a valence and arousal score that ranges between 1 and 5.  For instance, a sentence with high arousal and positive valence would be highly exciting and positive, or a low arousal and negative valence would be sad and calm.

2. Last.fm User 360k (Version 1.2) Logs \cite{celma2010music}: To personalize recommendations and extract behavioral trends, we use user profiles and play history data from Last.fm which contains interactions of users (like metadata, user history, and user activity logs) with the music tracks of Last.fm, a music streaming platform.

From this, we infer each user’s top artist and derive cross-product preference tables like $user \times emotion$ and $emotion \times artist$. These metrics serve as a memorization component in the wide component of the model and help rank song recommendations based on user-specific emotion-genre specifications.

3. Spotify Million Song Dataset (Lyrics Subset) \cite{spotifylyrics2023}: This dataset provides song metadata and lyrics for a subset of the Million Song Dataset. Each entry includes fields such as artist, song, and text (lyrics). We process the lyrics using the same transformer embedding and regression head pipeline to predict the VA values of each song. We use sentence binning for categorical wide features to ensure consistency with the EmoBank VA regressor. This results in a VA-annotated song database, which can be used to compute emotional distance between user input and candidate tracks during recommendation.

\subsection{Methodology}
We use a pretrained transformer called MiniLM or SBERT \cite{sbert}, evaluated on common Semantic Textual Similarity (STS) tasks and transfer learning tasks, to convert the user-inputted sentence into a 384-dimensional vector. Transformers were chosen because they can weigh important words regardless of their position, which enables contextual understanding. Since they are trained on millions of sentences, they are good at recognizing emotional and semantic features. Unlike CNNs (which are local) or LSTMs (which are sequential), transformers grasp global meaning efficiently. The system takes in the sentence from the user as input and generates a contextualized 384-dimensional embedding (\textit{deep feature}) of the sentence, representing semantic and emotional cues.


Simultaneously, we take the total number of words for each text from the 'text' column of the EmoBank dataset and split it into 7 buckets varying from 0 to 800, from which we create a one-hot encoded vector (\textit{wide feature}). Due to this, the model learns that texts of different lengths have different emotional expression patterns. 

Later, these two features are passed into a custom feed-forward neural network, which combines them to predict a two-dimensional VA vector. This network consists of two MLP branches, one of which processes the transformer output, and the other processes the wide features. These are then concatenated and passed through a final linear layer to regress the [V, A] values. The base transformer is LoRA-adapted (Low-Rank Adaptation of Attention) \cite{hu2021lora}, meaning only a small set of injected low-rank matrices inside the attention blocks are finetuned, while the rest of the transformer remains frozen due to computational limitations. This is also done for parameter-efficient training and avoids overfitting on the small EmoBank dataset. SmoothL1Loss is used instead of the standard MSELoss as it converged faster and gave a better loss. The ground truth VA values are standardized using StandardScale(). Additionally, training employs mixed precision and a warm-up linear learning rate scheduler to optimize performance.
\begin{figure}
    \centering
    \includegraphics[width=0.25\linewidth]{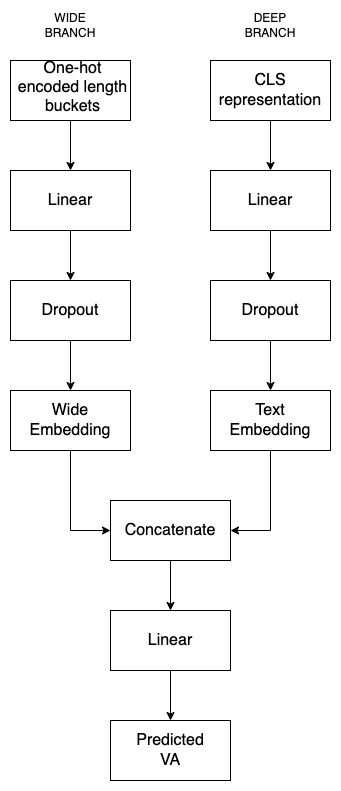}
    \caption{Wide and Deep architecture}
    \label{fig:enter-label}
\end{figure}

Once the model is trained on EmoBank, it predicts VA coordinates for each song in Spotify’s Million Song Dataset. Each song has hundreds of words, which our model might never have seen, which might result in poor generalization. To mitigate this, we pass chunks of songs to the model to predict VA scores and take the average of all the chunks to get the final VA score of that song.


Last.fm provides user metadata, including each user’s top artist and total listening activity, which we use to build personalized user profiles. Our system leverages two key features to enhance recommendation quality:

1. Top artist filtering : By aggregating their historical play counts, we identify the most played artist for each user. This artist is used for narrowing down the pool of candidate songs from the Spotify lyrics dataset to ensure relevance and familiarity.


2. Listening Frequency: To capture the user engagement level, the total number of plays is bucketed into multiple levels (such as low, medium, high, and super), which indirectly reflects the users' fondness for their top artist, and this is used for segmentation and analysis.


We then construct two memory tables using user listening histories combined with emotion predictions derived from lyrical content (V-pred and A-pred). These continuous scores are first grouped into discrete VA bins, representing coarse emotional categories:
\newline
1. $user \times emotion$ → how often the user listens to songs in a given emotional bucket. 
\newline
2. $emotion \times artist$ → how often an artist appears in songs of that emotional category.

To construct these memory tables, we first find the common artist names from the Last.fm user logs and Spotify song lyrics.
This is done via a string-based merge between the artist\_name field in the listening logs and the artist field in the lyrics dataset. 

One limitation of this method is that it assumes consistent formatting of artist names across datasets and may be affected by discrepancies such as case sensitivity, punctuation, or alternate naming conventions.

After matching, each play record is associated with the corresponding VA bin of the song, which enables us to aggregate plays per emotional category. 

The resulting memory tables are normalized row-wise to yield preference distributions: the first captures how much a user engages with each emotional region, while the second reflects how strongly an artist is associated with particular emotions. These distributions are later used to boost the recommendation score for songs that emotionally align with the user’s past preferences and favored artists.

Finally, during recommendation, the query sentence is encoded into a VA vector using the same trained transformer-based model used during training. A score is computed for each candidate song using:

\[\text{score} \;=\; -\left\lVert \mathbf{q}_{\text{VA}} - \mathbf{s}_{\text{VA}} \right\rVert_2
\;+\; \text{mem\_ue} \;+\; \text{mem\_ea}
\]

    \label{eq:godeq}

We calculate the Euclidean distance between the query and the song to measure its emotional similarity. A negative sign is used so that smaller distances — indicating better emotional alignment — contribute to the score positively. The term mem\_ue represents how often the user has listened to the songs in the same VA bin as the query, while the term mem\_ea indicates how often that song’s artist appears in the emotional bin, demonstrating emotional familiarity.

Before scoring, we filter the candidate pool using the user’s top artist to prioritize familiar content. If this yields no results, then the full song dataset is used. This way, our scoring approach makes sure that the recommended songs are emotionally aligned, historically relevant, and artistically familiar. Finally, the top k-songs with the highest scores are returned to the user as the final recommendations.




The model was then evaluated by giving an input manually.
E.g., “sitting in the grass and looking at mountains in pleasant weather”.
This sentence is encoded in a VA vector, and the system computes the Euclidean distance in VA space between the query and pre-encoded songs, filtering by the user’s favorite artist where possible (based on Last.fm play counts). A sample of the top-5 recommended songs for one such query is shown below in Table~\ref{tabvalar}.

\begin{table}[h!]
\caption{High valence and low arousal prediction}
    \centering
    \begin{tabular}{|c|c|c|c|}
        \hline
        Artist & Song & V-pred & A-pred \\
        \hline
        Kirk Franklin & Oh Happy Day & 3.481344 & 3.186183 \\
        
        Enya & Afer Ventus & 3.445730 & 3.186253 \\
        
        Israel & Lord You Are Good & 3.454684 & 3.199386 \\
        
        Enya & Busted & 2.273 & 3.426 \\
        
        Children & Alouette (english Translation) & 3.384859 & 3.118401 \\
        
        Faith Hill & Holly Jolly Christmas & 3.455301 & 3.205785 \\
        \hline
    \end{tabular}
    
    \vspace{0.5em}

    \label{tabvalar}
\end{table}


Looking at the results, we can infer that the recommended songs closely align with the emotional tone expressed in the query. The given sentence conveys a sense of calmness, peace, and moderate positivity. This conveys a medium valence (pleasantness) and low-to-moderate arousal (energy level).
\newline
A sentence like 'Crying and panting a lot' portrays low happiness and the word 'panting' portrays high energy. It resulted in songs which are sad yet energetic as shown in Table~\ref{tab:table2}.

\begin{table}[h]

\caption{Low valence and high arousal prediction for the sentence, "Crying and panting a lot"}
    \centering
    \begin{tabular}{|c|c|c|c|}
        \hline
        Artist & Song & V-pred & A-pred \\
        \hline
        Ramones & Cretin Family & 2.240889 & 3.470645 \\
        
        Nick Cave & The Hammer Song & 2.254322 & 3.485341 \\
        
        Keith Green & On The Road To Jericho & 2.270143 & 3.439531 \\
        
        Who & Melancholia & 2.270583 & 3.462609 \\
        
        Metallica & Hardwired & 2.266423 & 3.482848 \\
        \hline
    \end{tabular}
    \vspace{0.5em} 
    
    \label{tab:table2}
\end{table}

\begin{table}[h]

\caption{High valence and high arousal prediction for the sentence, "This is the best day of my life!"}
    \centering
    \begin{tabular}{|c|c|c|c|}
        \hline
        Artist & Song & V-pred & A-pred \\
        \hline
        Harry Connick, Jr. & S'wonderful & 3.969359 & 3.768258 \\
        
        Perry Como & Joy To The World! & 3.918838 & 3.804626 \\
        
         Judy Garland & I Am Loved & 4.030313 & 3.691707 \\
        
         Religious Music & Awesome God & 4.047324 & 3.667170 \\
        
        Hank Williams Jr. & I Really Like Girls & 3.854384 & 3.759417 \\
        \hline
    \end{tabular}
    \vspace{0.5em} 
    
    \label{dogdiedfreedfood}
\end{table}

\begin{table}[h]

\caption{Low valence and low arousal prediction for the sentence, "I feel numb, like everything has lost meaning"}
    \centering
    \begin{tabular}{|c|c|c|c|}
        \hline
        Artist & Song & V-pred & A-pred \\
        \hline
        Christy Moore & Anne Lovett & 2.428474 & 3.165729 \\
        
        Santana & She's Not There & 2.405259 & 3.251843 \\
        
        Yo Gotti & Sorry & 2.387484 & 3.261565 \\
        
        Opeth & Karma & 2.457449 & 3.178025 \\
        
         Primus & Fisticuffs & 2.439502 & 3.229409 \\
        \hline
    \end{tabular}
    \vspace{0.5em} 
    
    \label{coolbreeze}
\end{table}

\begin{table}[h!]

\caption{High valence and high arousal prediction for the sentence, "Having an ecstatic morning"}
    \centering
    \begin{tabular}{|c|c|c|c|}
        \hline
        Artist & Song & V-pred & A-pred \\
        \hline
        Patsy Cline & That's My Desire & 3.624 & 3.691 \\
        
        Kelly Clarkson & At Last & 3.604 & 3.700 \\
        
        Annie  & We Got Annie & 3.617 & 3.672 \\
        
        Matt Redman & The Happy Song & 3.631 & 3.676 \\
        
        Janis Joplin & Oh My Soul & 3.603 & 3.706 \\
        \hline
    \end{tabular}
    \vspace{0.5em} 
    
    \label{ecstaticmorn}
\end{table}
Tables~\ref{dogdiedfreedfood},~\ref{coolbreeze}, and~\ref{ecstaticmorn} show the outcomes of our experiments with numerous different sentences, which show the users' feelings with the projected VA values. These examples demonstrate that the model doesn't just recommend slow or happy songs, but it targets songs whose combined emotional signal (from lyrics) is near the user's current emotional expression in the VA space.

\subsection{\textbf{Training results analysis}}
Consistent improvement in both training and validation performance was seen in just 5 epochs. Training was done only for 5 epochs, and finetuning a small portion of the transformer (via LoRA adapters on query/value weights) is done, so the number of learnable parameters is small. This setup leads to fast convergence even with limited epochs, unlike full transformer finetuning, which may require 10-30 epochs. Initially, the model starts with a training loss of approximately 0.337, which drops to 0.239 by the final epoch. This shows the effective learning of patterns from the EmoBank text and wide features. The validation loss follows a more gradual decrease, stabilizing around 0.253-0.254, indicating that the model is generalizing well and not overfitting. The validation R² score improves steadily and plateaus around 0.427-0.428, suggesting that the model explains around 42.7 percent of the variance in the Valence-Arousal space on unseen data. This is a strong performance for subjective emotional prediction tasks.
\begin{figure}
    \centering
    \includegraphics[width=0.75\linewidth]{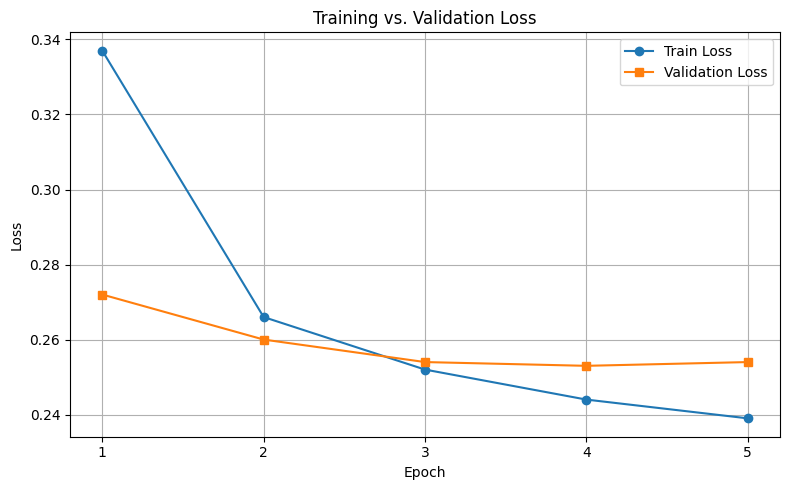}
    \caption{Training and Validation loss curves}
    \label{fig2}
\end{figure}

This convergence between training and validation losses suggests a well-regularized model. The use of LoRA adapters and SmoothL1 loss both contribute to stable training dynamics.

\section{Discussion}
We performed a few experiments with different hyperparameters and configurations, which are explained in this section.
To benchmark the effectiveness of our model, we also experimented with larger pre-trained models, including BERT-base and RoBERTa-base, finetuned in a similar regression setup. However, these models showed significantly poorer convergence. For these models, the training loss started at 140.00 and the validation loss at 70.00, which are orders of magnitude higher than our MiniLM-based system. While both models did eventually improve, they began overfitting rapidly - BERT after 17 epochs and RoBERTa \cite{roberta} after 15. In contrast, our system using MiniLM with LoRA \cite{hu2021lora} adapters converged within just 5 epochs and exhibited stable validation behavior with a lower final loss and better R².
\newline
\newline
This might be due to the following reasons:
\newline

1. MiniLM is a much smaller and computationally efficient model, optimized for sentence-level tasks. EmoBank \cite{buechel-hahn-2017-emobank} contains short text entries, making it more compatible with MiniLM’s pretraining objectives. In contrast, BERT and RoBERTa, with their larger parameter counts and broader pretraining on sentence-pair or masked language modeling tasks, require more data and careful regularization to avoid overfitting. This is something hard to control with smaller emotion datasets.

2. Our approach finetunes only a subset of attention weights using LoRA, greatly reducing the risk of overfitting while still adapting to the emotional domain.

We also compared loss functions and found that Smooth L1 loss consistently outperformed Mean Squared Error (MSE). MSE tended to produce unstable training and converged at a higher training loss, likely due to its sensitivity to outliers in VA scores. 
Smooth L1 (also known as Huber loss) combines the best of MSE and MAE by being quadratic for small errors and linear for large ones, making it more robust for noisy, subjective emotion annotations.

\section{Conclusion}

\begin{figure}[h!]
    \centering
    \includegraphics[width=1\linewidth]{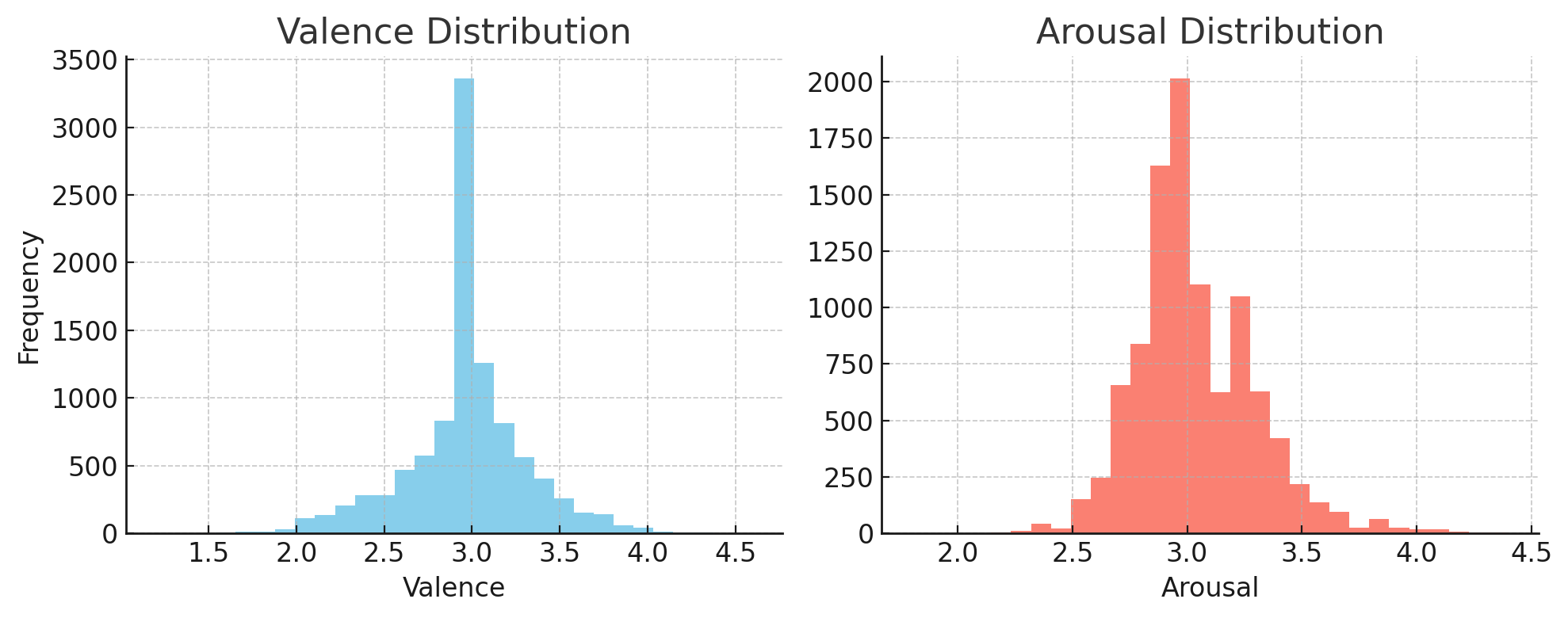}
    \caption{Valence and Arousal count for EmoBank dataset}
    \label{fig:enter-label-emo}
\end{figure}

\begin{figure}[h!]
    \centering
    \includegraphics[width=1\linewidth]{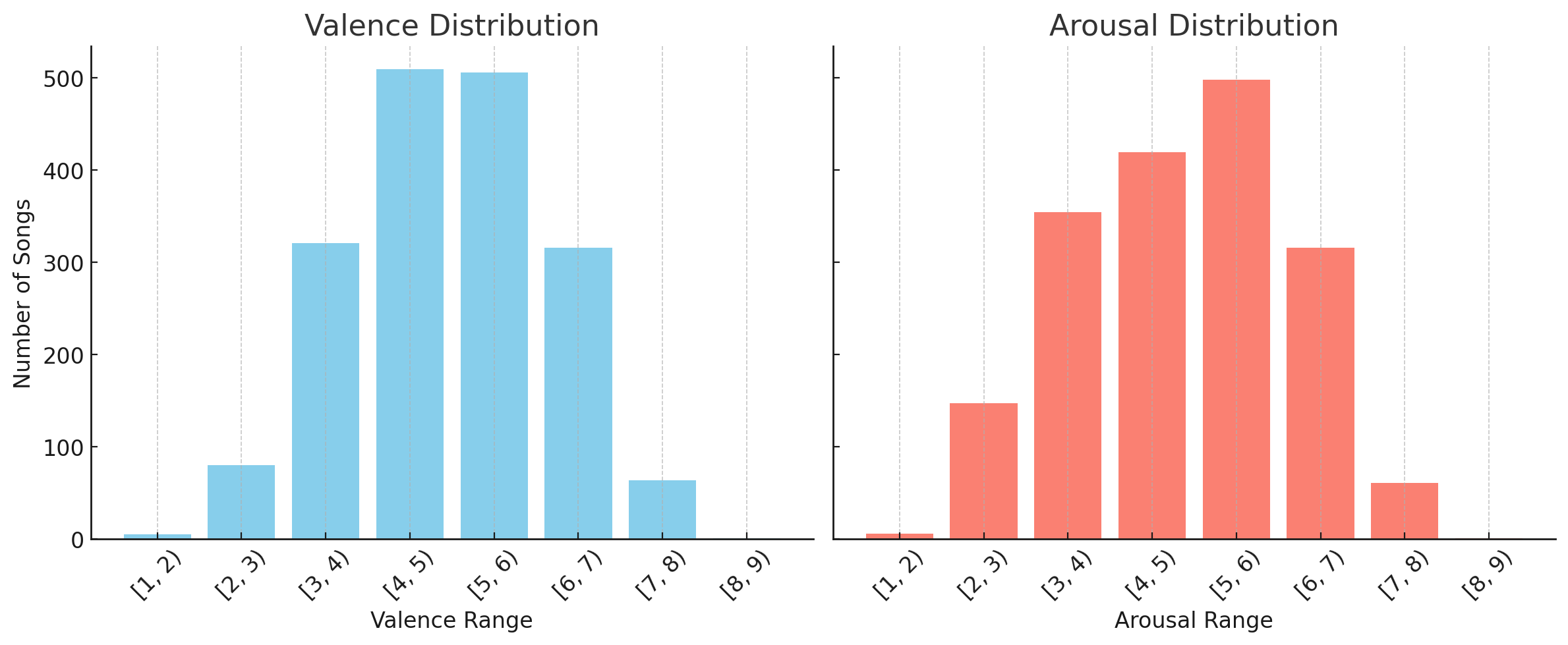}
    \caption{Valence and Arousal count for DEAM dataset}
    \label{fig:enter-label-deam}
\end{figure}

We introduced a novel methodology that involves moving beyond tags or discrete mood labels for music recommendation. We used  $emotion \times user$  and $emotion \times artist$ memory table learned from playcounts and combined VA distance with user preferences for a personalized scoring function and mood-based recommendation.
We also found that there are not many values that have a valence score less than 2 and greater than 4 (range 1-5). Similarly, there are very few values that have an arousal less than 2.5 and greater than 3.5 in the EmoBank dataset. There are only 6 entries in the valence extreme ranges ( less than 1.5 or greater than 4.5), and 0 entries in the arousal extremes (Figure~\ref{fig:enter-label-emo}). This shows that EmoBank has very few samples in the emotional extremes, which may limit the model’s ability to learn or predict strong emotions. We also checked the score values in another popular dataset called DEAM (Figure~\ref{fig:enter-label-deam}). It can be observed that there are minimal values on the extrema of this dataset. Many new datasets are available \cite{survey-on-datasets} in the field, which can be explored for this project.

We completed this project with limited data available on the internet and with limited computational power. We wanted to introduce the idea and prove that it works. Last.fm and Spotify only shared a few artists whose names matched well. With more robust datasets where artist names, users, play counts, genres, and their songs are in the same place, it would be much easier. 
\subsubsection{\discintname}
The authors have no competing interests to declare that are relevant to the content of this article.

\end{document}